\documentclass[aps,prl,twocolumn,showpacs,preprintnumbers,amsmath,amssymb]{revtex4}
\usepackage{graphicx}
\usepackage{dcolumn}% Align table columns on decimal point
\usepackage{bm}% bold math%
\newcommand{\futoi}[1]{\mbox{\boldmath$#1$}}

\begin{document}
\title{Evolution of the spin Hall effect in Pt nanowires: Size and temperature effects}

\author{Laurent Vila$^1$, Takashi Kimura$^{1,2}$, YoshiChika Otani$^{1,2}$}
\email{yotani@issp.u-tokyo.ac.jp}

\affiliation{%
$^1$ Institute for Solid State Physics, University of Tokyo, Kashiwanoha, Kashiwa, Chiba 277-8581, Japan \\
$^2$ Frontier Research System, RIKEN, Wako, Saitama 351-0198, Japan
}%

\date{\today}

\begin{abstract}
We have studied the evolution of the Spin Hall Effect in the regime 
where the material size responsible for the spin accumulation 
is either smaller or larger than the spin diffusion length. 
Lateral spin valve structures 
with Pt insertions were successfully used to measure the spin absorption 
efficiency as well as the spin accumulation in Pt induced through 
the spin Hall effect.  
Under a constant applied current the results show 
a decrease of the spin accumulation signal 
is more pronounced as the Pt thickness exceeds the spin diffusion length.  
This implies that the spin accumulation originates from 
bulk scattering inside the Pt wire and the spin diffusion length limits the SHE. 
We have also analyzed the temperature variation 
of the spin hall conductivity to identify the dominant scattering mechanism.  

\end{abstract}

                            
\maketitle

Recently a long-standing prediction of Spin Hall Effect (SHE) 
\cite{Dyankonov, Hirsch, Zhang} has been verified by means of both 
optical \cite{Kato, Wunderlisch, Sih, Stern} and magneto-transport 
\cite{Valenzuela, KimuraPRL, Saitoh} measurements.  
The SHE originates from the spin-orbit coupling, 
which relates the spin of an electron to its momentum, 
producing a spin current in the direction transverse to the flow of 
electrons and a spin accumulation at lateral sample boundaries. 
This is the direct SHE (DSHE) where unpolarized charge currents 
are converted into pure spin currents with zero net charge flow.  
There is also the inverse effect called inverse SHE (ISHE)
where a pure spin currents can be converted into charge currents by 
the spin-orbit interaction.  
The SHE is therefore regarded as a new tool for generating 
and detecting spin currents, crucial issues for future \emph{spintronics}, 
that in principle does not require 
ferromagnetic elements and/or external magnetic field.

The SHE induced spin accumulation in Semiconductors (SCs) have 
drawn much attention because of its compatibility 
with conventional CMOS technology. 
However, up to now SCs have exhibited very small SHE 
and no electrical detection has been reported yet.  
On the contrary Pt metal has been successfully 
used to detect the SHE even at room temperature, exhibiting the largest spin 
Hall conductivity reported so far \cite{KimuraPRL}. 
The possible origins of the SHE can be classified 
in two categories, intrinsic and extrinsic, 
depending on the dominant influence of either 
band structure or impurities \cite{Revue}. 
The SHE is also expected to be strongly geometry dependent. 
In metals like Pt, the SHE may be mainly attributable to 
extrinsic mechanisms such as the side jump and the skew scattering, 
which are responsible for the anomalous Hall effect in ferromagnets. 
The different sign of the side jump or the skew scattering angle 
for the two spin channels results in the transverse spin current 
and spin accumulation at lateral boundaries.  
One should also remark here that recent theoretical analysis 
based on the first principle band calculation strongly suggests that 
the origin of the large SHE in Pt is of intrinsic nature \cite{Guo}. 
Thus it is important to perform systematically experiments 
in order to better understand the origin of the SHE.

The first electrical detection scheme, originally proposed 
by Takahashi and Maekawa \cite{Takahashi}, 
was successfully applied for alminum by Valenzuela \emph{et al} 
\cite{Valenzuela}.   
But this method is not suitable for the material with a nanometer scale 
spin diffusion length $l_{sf}$ due to strong spin-orbit interaction.  
Therefore in a previous study\cite{KimuraPRL} 
we have demonstrated a new device design 
where the Cu wire transfers accumulated spins 
at the interface between ferromagnetic detector (or injector) and Pt  
with negligible relaxation for both DSHE and ISHE measurements.   
In the previous study, we assumed that the induced spin current 
at the Cu/Pt interface was completely absorbed by the Pt wire 
because of its small spin resistance.  
However, the absorption efficiency of the spin current may depend on 
the device geometry and the temperature.  
Therefore, we further modified the previous design to a lateral spin valve (LSV) 
structure with a Pt insertion, 
which enables us to determine explicitly the magnitude of the absorbed spin current. 
Thereby the spin Hall conductivity for the Pt wire can be well evaluated.

\begin{figure}
\includegraphics[width=8.66cm]{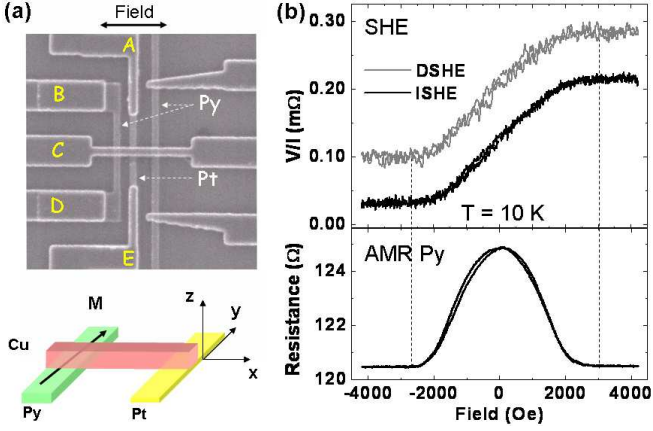}% Here is how to import EPS art
\caption{\label{fig:Fig1} (color online) 
(a) SEM image of the typical device for SHE measurements and an illustration of the device. 
(b) Direct and Inverse SHE (DSHE and ISHE) recorded at T = 10 K using a device with $t_{Pt}$ = 20 nm, 
altogether with the AMR from the Py wire measured on the same condition.}
\end{figure}

Our device is based on the LSV geometry \cite{Jedema, JedemaPRB,KimuraPRB}  
where a Pt wire is inserted, which can be either a spin current source for DSHE 
or a spin current absorber for ISHE experiments.  
The devices are prepared by means of conventional e-beam lithography, 
electron gun evaporation (base pressure $\sim$ $10^{-8}$ Torr) 
and a lift-off process. 
The scanning electron image of a typical device is 
shown in Fig.~\ref{fig:Fig1}.(a). 
A Pt wire 100 nm wide ($w_{\rm Pt}$) with a thickness $t_{Pt}$ varied 
from 5 to 20 nm is in ohmic contact with a Cu wire running 
between two Permalloy (Py) wires 100 nm in width and 30 nm in thickness.  
The dimensions of the Cu wire are 160 nm in width and 100 nm in thickness.  
Prior to Cu evaporation a careful Ar Ion Beam Etching is carried out 
for cleaning surface to obtain 
highly transparent ohmic contacts \cite{KimuraPRB}.  
The measurements are carried out by using ac lock-in amplifier 
and an He flow cryostat. 
The magnetic fields between $-$ 4 kOe and $+$ 4 kOe are
applied in the plane of the substrate. 

The charge to spin conversion induced by SHE obeys 
a relation given by the vector product, 
$\futoi{J_s} \propto \futoi{S}$ $\times$ $\futoi{J_c}$ with $\futoi{J_{s(c)}}$ 
the spin (charge) current density and $\futoi{S}$ the spin orientation.  
Similarly the spin to charge conversion obeys the reciprocal relation, 
$\futoi{J_c} \propto \futoi{S} \times \futoi{J_s}$.  
In the case of DSHE, the spins of the unpolarized charge current in the Pt wire 
are therefore aligned along the $x$-axis, 
accumulating in the vicinity of the top or bottom surface
according to the spin orientation.  
The accumulated spins are transferred through the Cu wire 
to the left Py spin detector in Fig. 1 (a).  
The magnitude of the spin accumulation is measured as the spin signal 
$R_S = V_{\rm BC}/I$ where $V_{\rm BC}$ is the voltage between 
the contact leads B and C and $I$ ($ = 80 \mu$A) is the current flowing in the Pt wire 
through contact leads A and E. 
When the magnetic field is applied along $x$ axis, 
the spin signal $R_S$ exhibits a linear increase and 
gets saturated above the saturation field, 
reflecting a hard axis magnetization process of Py wire.  
This magnetization process is separately confirmed by the anisotropic 
magnetoresistance measurement as in the bottom of Fig. 1 (b).
The overall change $\Delta R_S$ in the signal is about 0.18 m$\Omega$.

In the case of the ISHE curve in Fig.~\ref{fig:Fig1}.(b), 
the spin current is injected by flowing a current of 240 $\mu$A between B and C. 
The voltage is induced between E and A as a result of the spin to charge conversion 
since the spin current is preferentially absorbed in the Pt wire along the $z$-axis.  
From the field variation, the overall change in charge accumulation signal 
$\Delta R_{SHE}$ is also determined to be 0.18 m$\Omega$ which is equal to $\Delta R_S$ in the DSHE, 
indicating the reciprocity of the behavior.  
The angular variation of the spin and charge accumulation signals 
measured with rotating the applied field direction 
assure the relation among the vectors $\futoi{J_s}$, $\futoi{J_c}$ and $\futoi{S}$.

\begin{figure}
\includegraphics[width=8cm]{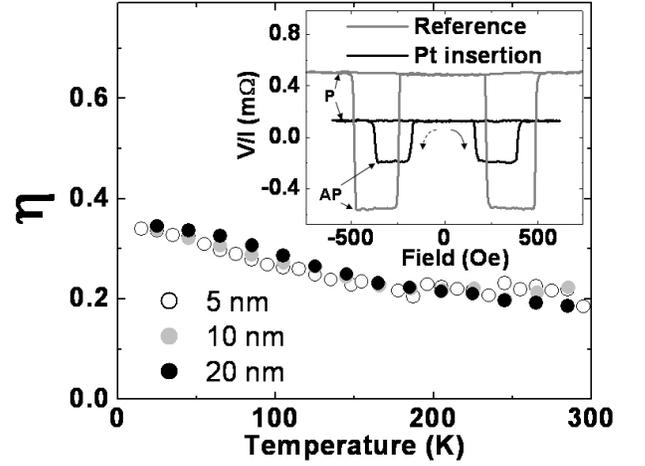}% Here is how to import EPS art
\caption{\label{fig:Fig2}
Temperature evolution of the spin absorption for Pt wires 
with $t_{Pt}$=5, 10 and 20 nm and $w_{Pt}$=100 nm. 
The inset shows nonlocal spin valve signals measured at 5 K for LSVs with 
and without Pt insertion ($t_{Pt}$ = 5 nm) clearly indicating 
parallel (P) and anti-parallel (AP) states.}
\end{figure}

The efficiency of the spin absorption is studied 
by non local spin valve (NLSV) measurements detailed in \cite{KimuraPRB}. 
For this measurement the magnetic field is applied along the Py wires.  
For all devices the center to center distance $d$ between the Py wires 
is fixed to 800 nm. 
The inset in Fig.~\ref{fig:Fig2} compares the NLSV signals 
measured at 5 K for the devices with and without a Pt insertion 
between the two Py electrodes.   
The NLSV signal for $I \sim$ 240 $\mu$A clearly decreases 
once the Pt is inserted, indicating the spin current absorption 
into the Pt wire, in good agreement with previous experiments \cite{KimuraPRB}. 
For all the devices the NLSV signal is almost a factor of 5 decreased 
by the Pt insertion from 5 K to room temperature (RT). 
We now define the ratio $\eta$, which is the ratio of the spin signal with and without a Pt insertion, 
$\Delta R_{\rm SV}^{\rm with}$ and $\Delta R_{\rm SV}^{\rm ref}$, repspectively.  
By solving the one dimensional spin diffusion model with transparent interfaces, 
$\eta$ can be calculated as 
\begin{equation}
\eta \equiv \frac{\Delta R_{\rm SV}^{\rm with}}{\Delta R_{\rm SV}^{\rm ref}}
\approx \frac{\frac{R_{\rm Pt}}{R_{\rm Cu}} \sinh \left( d/{l_{sf}^{\rm Cu}} \right)}
{\cosh \left( {d}/{l_{sf}^{\rm Cu}} \right) + 
\frac{R_{\rm Pt}}{R_{\rm Cu}} \sinh \left( {d}/{l_{sf}^{\rm Cu}} \right)-1}.
\end{equation}
Here, $R_{\rm Pt}$ and $R_{\rm Cu}$ are the spin resistances for the Pt and Cu wires, respectively.  
$l_{sf}^{\rm Cu}$ is the spin diffusion length for the Cu wire.

Interesting is that the ratio $\eta$ varies from 0.35 at 5K to 0.2 at RT irrespective of the Pt thickness.  
This implies that the spin absorption into the Pt wire 
takes place only at the top interface between Pt and Cu wires. 
The reason for this may be due to the anisotropic Ar milling process which left side surfaces of the Pt insertion uncleaned.
The $\eta$ deduced from the experimental results 
yields the spin resistance of the Pt wire as 0.22 $\Omega$ both at RT and 5 K .  
Since the spin resistance of the Pt wire is given by $2l_{sf}^{\rm Pt}/(\sigma_{\rm Pt} S)$, 
where $\sigma_{\rm Pt}$ and $S$ are the conductivity for the Pt and the effective cross section 
for the spin current in the Pt wire, 
$l^{\rm Pt}_{sf}$ can be estimated as 14 nm at 5 K and 10 nm at RT,
in good agreement with the value of 14 $\pm$ 6 nm at 4 K reported by Kurt \emph{et al} \cite{Kurt}.

\begin{figure}
\includegraphics[width=8.66cm]{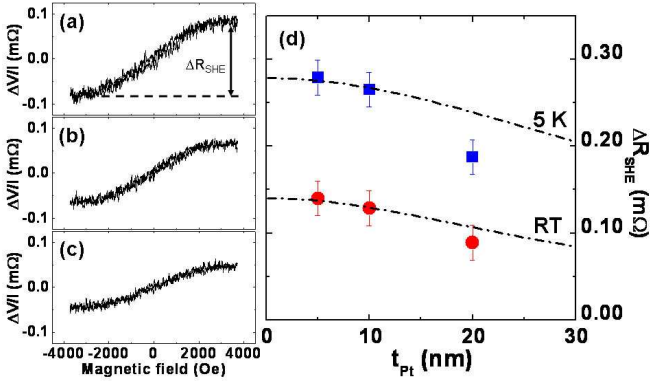}% Here is how to import EPS art
\caption{\label{fig:Fig3} (color online) 
Inverse spin Hall effect measured at RT for (a) $t_{\rm Pt}$ = 5 nm, (b) 10 nm and (c) 20 nm. 
The distance $d$ is of the order of 350 nm. (d) $\Delta R_{SHE}$ 
as a function of $t_{Pt}$ at 5 K and RT for $d =$ 350 nm. 
The dashed lines are the calculated evolution of $\Delta R_{SHE}$ using Eq.\ref{eq:eq3}.}
\end{figure}

Figures ~\ref{fig:Fig3} (a)-(c) represent the field dependences 
of the ISHE with different Pt thicknesses of 5, 10 and 20 nm measured at RT.  
Here, the distance $d$ is 350 $\pm 20$ nm.  
We have studied about 20 devices devices of $t_{\rm Pt}$ = 5, 10, and 20 nm, 
with different values of $d$ varied from 250 to 500 nm 
to evaluate the thickness dependence of the SH signal.  
All the devices exhibit a gradual decrease of DSHE and ISHE with $d$ 
proportional to $\sinh^{-1}(d/l^{\rm Cu}_{sf})$ due to the spin relaxation in the Cu wire.  
Interestingly, a clear systematic decrease in $\Delta R_{SHE}$ with $t_{Pt}$ 
has been observed in the temperature range from 5 K to RT.  
Figure \ref{fig:Fig3}.(d) summarizes the $\Delta R_{SHE}$ values at RT and 5 K 
obtained for the distance $d$ of 350 nm.  
There is a very small decrease in $\Delta R_{\rm SHE}$ while the thickness $t_{\rm Pt}$ is below 10 nm.
On the contrary there is a more pronounced decrease when the thickness is increased from 10 to 20 nm.  
This can be understood as follows.  
The spin accumulation at the top surface of the Pt due to the DSHE 
effectively takes place in the Pt while the thickness $t_{\rm Pt} < l^{\rm Pt}_{sf}$. 
Once $t_{\rm Pt}$ exceeds $l_{sf}^{\rm Pt}$, 
the spin accumulation should not increase under the condition of the constant current.
For the ISHE, the vertical penetration of the spin current from Cu into Pt is limited by $l^{\rm Pt}_{sf}$, 
leading to the same $\Delta R_{\rm SHE}$ as $\Delta R_{S}$.  
For more quantitative analyses, we calculate the thickness dependence of 
the spin accumulation  $\Delta V_{S}$ due to DSHE 
at the top surface of the Pt wire.  
Since the spin relaxation in the Cu wire does not depend on $t_{\rm Pt}$, 
the thickness dependence of $\Delta R_{S}$ or $\Delta R_{\rm SHE}$ 
is proportional to $\Delta V_{S}$.  
Using the homogeneous spin Hall conductivity $\sigma_{\rm SHE}$ in the Pt wire 
with open circuit condition, 
$\Delta V_{S}$ can be calculated as  
\begin{equation}
\Delta V_{S} = \frac{\sigma_{\rm SHE} I_{C}}{\sigma_{\rm Pt}^2 }
\frac{l_{sf}^{\rm Pt} }{w_{\rm Pt}t_{\rm Pt}}
\frac{\exp (t_{\rm Pt}/l_{sf}^{\rm Pt})-1}{\exp (t_{\rm Pt}/l_{sf}^{\rm Pt})+1}, 
\label{eq:eq3}
\end{equation}  
where $I_{C}$ is the excitation current flowing in the Pt wire.  
As shown in Figure 3(d), Eq.\ (2) with $l_{sf}^{\rm Pt} = 10$ nm at RT and 14 nm at 5 K, 
which are deduced from the absorption experiment mentioned previously, 
fairly well reproduce the thickness evolution of $\Delta R_{\rm SHE}$. The data can be well reproduced when slightly reducing $l_{sf}^{\rm Pt}$ to 7 nm and 8 nm for RT and 5 K respectively. 
These features strongly support the bulk origin of the DSHE and ISHE.

Finally, we discuss the temperature dependence of the spin Hall conductivity.  
The spin Hall conductivity $\sigma_{\rm SHE}$ can be calculated as 
\begin{equation}
\sigma_{\rm SHE} = w_{\rm Pt} \sigma_{\rm Pt}^2 \frac{I_c \Delta R_{SHE}}{I_s}
\label{eq:eq2}
\end{equation}
with the parameters in \cite{Data} to estimate the ratio $I_c / I_s$ of charge to spin currents. 
The obtained value of $\sigma_{\rm SHE}$ is 330 S/cm at RT, 
which is larger than that in the previous experiment.  
This is because the assumption of the complete spin current absorption into the Pt wire 
in the previous analysis led to underestimate the spin Hall conductivity.

\begin{figure}
\includegraphics[width=8cm]{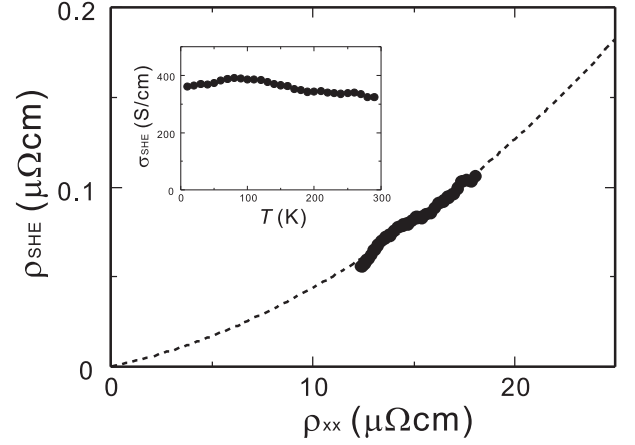}% Here is how to import EPS art
\caption{\label{fig:Fig4} 
Evaluated $\rho_{SHE}$ as a function of $\rho_{Pt}$ for the device in 
Fig.~\ref{fig:Fig3}.(a) with $t_{Pt}$ = 5 nm and $d$ = 325 nm. 
Fitting to the polynomial yields 
$\rho_{SHE} = 2.5 \times 10^{-3} \rho_{Pt} {\rm [\mu \Omega cm]} 
+ 1.9 \times 10^{-4} (\rho_{Pt} {\rm [\mu \Omega cm]})^2$. 
The inset represents the calculated $\sigma_{SHE}$ as a function of temperature using Eq.~\ref{eq:eq2}.}
\end{figure}

Using Eq.\ 3 with the temperature dependences of $\Delta R_{\rm SHE}$ 
and the spin resistances for Pt, Cu and Py, we can compute $\sigma_{\rm SHE}$. 
As shown in the inset of Fig.\ 4, 
the temperature dependence exhibits a very small variation of $\sigma_{\rm SHE}$ about 350 S/cm.  
%The spin Hall angle $\alpha_{SHE} = \sigma_{SHE}/\sigma_{Pt}$ 
%is then about 1.12 $10^{-2}$ and 1.42 $10^{-2}$ at 5 K and RT 
%respectively, wcgucg are the largest values reported so far. 
It should be noted that the change in $\sigma_{\rm SHE}$ is much smaller 
than that in \cite{Guo} obtained from band structure calculation. 
Our values of $\sigma_{SHE}$ are smaller at low temperatures but 
larger at RT (2000 S/cm and 240 S/cm in \cite{Guo} for low temperature and RT respectively). 
This difference could be 
due to the dominant contribution of impurities compared to the band structure considerations in the Pt wires. 
In Fig.~\ref{fig:Fig4} 
the SH resistivity ($\rho_{\rm SHE} \approx \frac{\sigma_{\rm SHE}}{\sigma_{\rm Pt}^2}$) is 
likely to evolve in a quadratic form with the Pt resistivity $\rho_{\rm Pt}$ 
with $\rho_{Pt}$ in this temperature range. 
This is expected evolution for side jump origin of the SHE.

In conclusion we have studied the evolution of the spin Hall effect in Pt wires 
in the regime where the size responsible for the spin accumulation 
is either smaller or larger than the spin diffusion length.  
Under the constant current the pronounced decrease in the SHE for 
$t_{Pt} > l_{sf}^{\rm Pt}$ demonstrates that the SHE is induced by bulk scattering inside the Pt wire.  
The analyses on the spin Hall conductivity of the Pt wires suggest a side jump origin of the SHE.  
These results will provide important information in determining the optimum geometry for the SHE. 

We thank Profs. Y. Iye and S. Katsumoto of ISSP, Univ. of Tokyo for the use of the lithography facilities.


\begin{thebibliography}{99}
\bibitem{Data} From independant measurements $l_{sf}^{Cu}$, $R_N$, $R_F$ and $p$ were estimated to be respectively 500 nm, 1.3 $\Omega$, 0.09 $\Omega$, 0.2 at RT and 1500 nm, 1.7 $\Omega$, 0.13 $\Omega$, 0.35 at 5 K.
\end{thebibliography}
\end{document}